\documentclass[aip,jcp,reprint]{revtex4-1}
\usepackage{graphicx}
\usepackage{amsmath,amssymb}
\usepackage{upgreek}
\usepackage{enumitem}

\usepackage{color}

\newcommand{\kB}{k_{\mathrm{B}}}
\newcommand{\kT}{\kB T}

\newcommand{\Pe}{\mathrm{Pe}}

\newcommand{\Pefilm}{\Pe_{\mathrm{film}}}

\newcommand{\vev}{v_{\mathrm{ev}}}
\newcommand{\vjam}{v_{\mathrm{jam}}}
\newcommand{\phijam}{\phi_{\mathrm{jam}}}
\newcommand{\tstarjam}{t^*_{\mathrm{jam}}}
\newcommand{\tstarend}{t^*_{\mathrm{end}}}

\newcommand{\phiinit}{\phi_{0}}

\newcommand{\Dsmall}{D_{\mathrm{small}}}
\newcommand{\Rbig}{R_{\mathrm{big}}}
\newcommand{\Rsmall}{R_{\mathrm{small}}}

\newcommand{\vbig}{v_{\mathrm{big}}}
\newcommand{\zint}{z_{\mathrm{int}}}
\newcommand{\zjam}{z_{\mathrm{jam}}}

\newcommand{\tstar}{t^*}

\newcommand{\phis}{\phi_{\mathrm{small}}}
\newcommand{\rhos}{\rho_{\mathrm{small}}}

\newcommand{\Peclet}{P\'eclet}

\newcommand{\latin}[1]{{\itshape #1}}

\newcommand{\ie}{\latin{i.\,e.}}
\newcommand{\etal}{\latin{et al.}}

\DeclareMathOperator\erfc{erfc}

\newcommand{\Eqref}[1]{Eq.~\eqref{#1}}

\newcommand{\Figref}[1]{Fig.~\ref{#1}}

\begin{document}

\title{Stratification of mixtures in evaporating liquid films occurs
only for a range of volume fractions
of the smaller component}

\author{Richard P. Sear}

\affiliation{Department of Physics, University of Surrey, Guildford,
  GU2 7XH, UK}

\email{r.sear@surrey.ac.uk}

\begin{abstract}
I model the drying of a liquid film containing small and big colloid particles. Fortini et al. [A. Fortini \etal, {Phys.\ Rev.\ Lett.} {\bf 116}, 118301 (2016)] studied these films with both computer simulation and experiment. They found that at the end of drying the mixture had stratified with a layer of the smaller particles on top of the big particles. I develop a simple model for this process. The model has two ingredients: arrest of the diffusion of the particles at high density, and diffusiophoretic motion of the big particles due to gradients in the volume fraction of the small particles. The model predicts that stratification only occurs over a range of initial
volume fractions of the smaller colloidal species. Above and below this range the downward diffusiophoretic motion of the big particles is too slow to remove the big particles from the top of the film, and so there is no stratification. In agreement with earlier work, the model also predicts that large P\'{e}clet numbers for drying are needed to see stratification.
\end{abstract}

\maketitle

\section{Introduction}

Fortini {\it et al.} \cite{fortini16} studied
the drying of a liquid film containing a mixture of large and small
colloidal particles. They found spontaneous stratification in the final
dry film, with a layer enriched in the small particles on top of a layer
with the larger particles. This is a novel out-of-equilibrium
self-organisation mechanism, and potentially has applications.
For example, by using small and large particles with different
properties, the properties
of the top and bottom surfaces of the final film, could be independently
controlled.

Not all mixtures stratify \cite{schulz18}. For example,
both Mart\'{i}n-Fabiani {\it et al.} \cite{martin16},
and Makepeace {\it et al.} \cite{makepeace17} studied systems with
high initial volume fractions, and found
no stratification.
Motivated by this observation, I develop
a simple model to predict which mixtures of small and large particles
will stratify, and which will not.
I combine earlier work by Sear and Warren \cite{sear17} on modelling
dilute mixtures in drying films, with Okuzono {\it et al.} \cite{okuzono06}'s
work on dynamical arrest in drying solutions of polymers.
Okusono {\it et al.} \cite{okuzono06} developed a simple model for
a system where the dynamics arrests at high concentrations.
My combined model makes simple analytical predictions for which
films should stratify, and which should remain homogeneous.

I consider a thin liquid film of initial
height $H$ that
contains a colloidal dispersion.
This dispersion is a mixture of colloidal particles
with a small radius, $\Rsmall$, and particles
with the much larger radius, $\Rbig$.
The liquid
is volatile, and as it evaporates
the liquid/air interface
descends at the velocity $\vev$.
My model includes the effect of solvent flow, which
Sear and Warren \cite{sear17} have shown to be important,
but it has limitations. I can only consider the limit
of a large size ratio, $\Rbig/\Rsmall\gg 1$, and dilute concentrations
of large particles.
See either Keddie and Routh's book \cite{Keddie:2010ta} or Routh's review \cite{routh13review}
for an introduction
to drying films of colloidal particles, and their applications.

If evaporation is slow, then I assume that the colloidal
mixture will slowly compress until it jams or crystallises,
at a volume fraction of around 0.64 \cite{torquato00,meeker97,auer01,filion10}.
However, fast evaporation velocities
cause the particles to accumulate
immediately beneath the descending
interface \cite{schulz18,routh13review,Keddie:2010ta,Routh:2004jz,trueman12}.
Here, for the small particles,
\lq fast\rq~means a film
evaporation \Peclet~number
larger than one.
The \Peclet~number for the smaller species is defined by
\begin{equation}
\Pefilm=\frac{\vev H}{\Dsmall}
\end{equation}
where $\Dsmall$ is the diffusion constant
of the smaller species.

Drying suspensions of colloidal
particles have been studied extensively at large P\'{e}clet numbers
\cite{Keddie:2010ta,routh13review},
and the accumulation of particles below the descending water/air interface
is well understood. As large
and small particles have different
P\'{e}clet numbers, rapid drying
always creates differential accumulation
in mixtures of particles. A number
of studies \cite{luo08,trueman12,trueman12langmuir,Nikiforow:2010bi,schulz18,utgenannt16} prior
to that of Fortini {\it et al.} \cite{fortini16}
considered this differential accumulation and the resulting stratification.

The innovation
of Fortini and co-workers \cite{fortini16} was to show that stratification can be obtained
by diffusiophoretic motion of the larger
species.
Diffusiophoretic motion is, by definition,
motion of one species due to a gradient in concentration
of another species \cite{And89,And86,ruckenstein81,Bra11,derjaguin47,SUS+16,PAN+15,shin17,bocquet10}.
Here, diffusiophoretic
motion is motion of the large colloidal particles
in a concentration gradient of the small
particles. This concentration gradient
is produced by the descending water/air
interface.
Since the work of Fortini and co-workers, there have been a number
of computer simulation, modelling and experimental
studies of the drying of liquid films that contain
mixtures of small and big particles. These studies have all  observed
stratification
\cite{fortini17,sear17,HNP17a,HNP17b,zhou17,makepeace17,martin16,sear17,liu18}. The results
of recent experimental work is mostly consistent with diffusiophoretic
driven stratification \cite{martin16,makepeace17,liu18}, although the
simple models used in theory and simulation clearly do not capture all the
behaviour seen in experiment.

\begin{figure}[tbh!]
  \includegraphics[width=5.5cm]{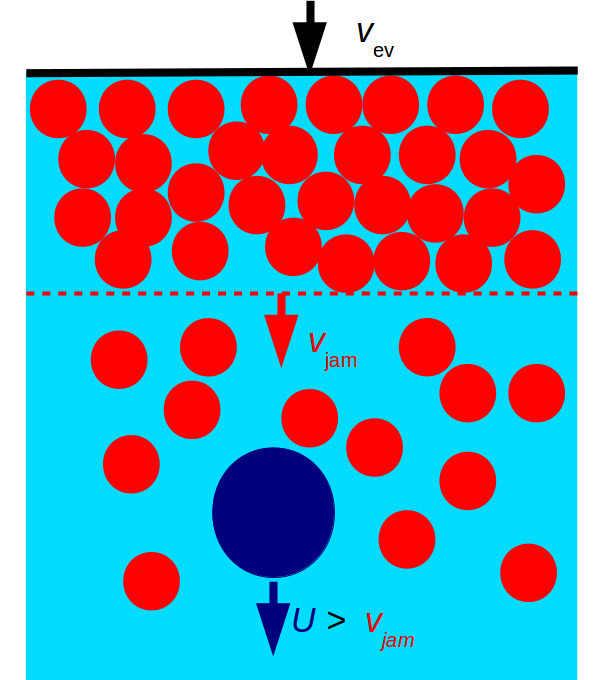}
\caption{Schematic illustrating what
needs to happen for stratification to occur.
As the water/air interface (black line) descends
at speed $\vev$, the small particles (red) accumulate
beneath it until they are so dense that they jam.
A growing jammed layer of small particles
then descends at speed $\vjam$. In front
of this jammed layer there is concentration
gradient that drives diffusiophoretic
motion of big particles (dark blue) at speed $U$.
If $U>\vjam$ then big particles are
excluded from the jammed layer of small
particles, and there is stratification.
\label{filmjam_schem}}
\end{figure}

Within the model studied here, stratification occurs when
the small particles accumulate, and then jam at high densities, under
such conditions that this jammed layer of small particles
excludes the big particles. This exclusion happens when
the downward diffusiophoretic motion of the big particles is
faster than the downward advance of the jammed layer of small particles.
I have illustrated this in \Figref{filmjam_schem}.
As the diffusiophoretic velocity is proportional to the gradient
in concentration of the small particles, this is equivalent
to saying that stratification requires large enough
concentration gradients below the jammed layer of small particles.

In the next section I describe
my adaptation
of Okuzono {\it et al.}'s model to describe the behaviour of the small colloidal particles.
In the third and fourth sections, I derive
expressions for the onset of jamming,
and for diffusiophoresis, respectively.
Results are in the fifth section,
while the sixth section is a conclusion.

\section{Okuzono {\it et al.}'s model applied to a one-component colloidal dispersion
in a drying film}

As a colloidal dispersion of hard
spheres is compressed to higher and
higher concentrations, the viscosity
increases, and the diffusion of the particles slows \cite{meeker97}.
Then one of two things happen: either the system
crystallises \cite{auer01,auer03,sanz11,filion10},
at which point the dynamics arrest,
or the volume fraction reaches values around 0.64 \cite{torquato00,liu98,meeker97}, at which point the system
is a glass, because the particles have been pushed into contact and
so their dynamics are again arrested.
Here, for simplicity I follow Okuzono {\it et al.} and assume
that the dynamics arrests and the system becomes a glass,
at a threshold density. I set the threshold volume fraction to be
$\phijam=0.64$, and refer to it as jamming.
When the particles are jammed I assume that the
descending water interface cannot compress them further.

In Okuzono {\it et al.}'s \cite{okuzono06}  model their polymer is an ideal solution
up to a gelling concentration, at which
point it becomes solid. They used
this model to understand \lq skin\rq~formation
in drying films of polymer solutions. This skin
is a gelled layer that forms at the top of the film,
where the concentration is highest.
The film is assumed to be infinite and uniform
in the $xy$ plane, with the water/air
interface moving down along the $z$ axis.

In my colloidal version
of Okuzono {\it et al.}'s model, the small colloid is a
diffusing ideal solution
when its local volume fraction $\phis(z)<\phijam$,
and is an incompressible solid
at $\phis(z)=\phijam$.
Therefore, the volume fraction
profile of the small colloid $\phis(z,t)$ obeys
the diffusion equation
\begin{equation}
\frac{\partial \phis}{\partial t}=\frac{\partial}{\partial z}
\left(D(\phis)\frac{\partial \phis}{\partial z}\right)
\label{diff_film}
\end{equation}
with
\begin{equation}
D(\phis)=\left\{\begin{array}{lc}
\Dsmall & \phis<\phijam \\
D_{\mathrm{skin}}\to\infty & \phis>\phijam
\end{array}\right.
\end{equation}
The large $D_{\mathrm{skin}}$ ensures that the \lq skin\rq~layer has
a uniform volume fraction equal to $\phijam$, while the constant
diffusion constant $\Dsmall$ below $\phijam$ means
that there \Eqref{diff_film} reduces to the diffusion equation for an ideal gas.

The boundary conditions are as follows. We have two walls,
at the top and bottom. The bottom wall is fixed at $z=0$, and
models the substrate the film is on.
The boundary condition at the bottom wall is zero flux.

The top
wall is the water/air interface. This interface starts at $\zint(t=0)=H$
and then descends at the fixed evaporation speed $\vev$.
The position of the interface
at time $t$ is given by
\begin{equation}
\zint(t)=H-\vev t=H(1-\tstar)H
\end{equation}
where we have defined the reduced time
\begin{equation}
\tstar=\frac{\vev t}{H}\quad ({}\le 1)
\end{equation}
The boundary condition at the descending top interface  is
again zero flux.
The final boundary condition is an initial condition, i.e., it is the initial state
of the state. At $t=0$, the small colloid is unifomly distributed
with a constant volume fraction $\phiinit$.

\begin{figure}
  \includegraphics[width=9cm]{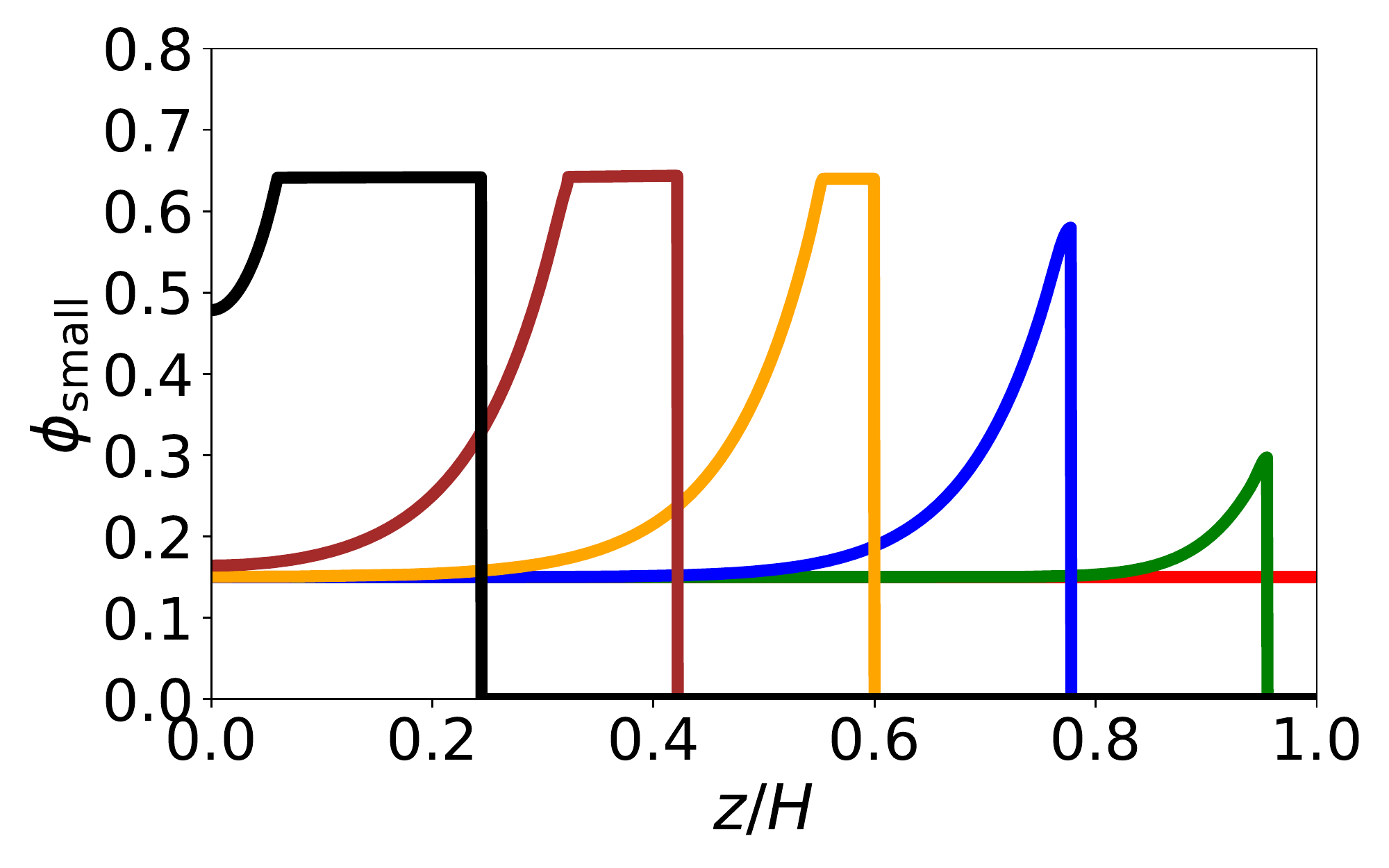}
\caption{Plots of the volume fraction as a function of height, for a
single-component dispersion of the small colloid.
The profiles are at times
$t^*=0$ (red), $0.044$ (green),
$0.22$ (blue), $0.40$ (yellow), $0.58$ (brown)
and $0.76$ (black).
$\Pefilm=10$, the initial
concentration $\phiinit=0.15$, and $\phijam=0.64$.
Profiles are obtained by numerically solving the diffusion PDE,
with $D_{\mathrm{skin}}/\Dsmall=1000$.
\label{phi_profiles}}
\end{figure}

\subsection{Example results for accumulation and jamming during drying}

In \Figref{phi_profiles}, I have plotted concentration
profiles at a number of different times during drying.
This is for a film with $\Pefilm=10\gg 1$.
As the water/air interface descends, the small
particles accumulate immediately below this interface.
During drying, the concentration will reach
$\phijam$, and this occurs first where the
concentration is largest, which is at the descending interface.
So a jammed layer starts at the top interface and grows
in thickness during drying. As it does so it is pushed down
until it reaches the bottom, at which time the dynamics in our simple
model stops.

We can compare the profiles of our simple model, which
is an ideal solution
up the jamming concentration, with the results of computer
simulations \cite{fortini16,HNP17a,HNP17b}
and density-functional theory \cite{HNP17a,HNP17b},
which include excluded-volume
interactions at all concentrations.
We note that our model underestimates the width of
the accumulation zone, compare our \Figref{phi_profiles},
with Fig.~2 of Fortini {\it et al.}\cite{fortini16},
and with Fig.~4 of Howard {\it et al.}\cite{HNP17a}.
As we will see in the next section, within our model the accumulation
zone has a width of $\Dsmall/\vev$. Whereas when interactions
are taken into account the profiles are a few times wider than this.

\section{Approximate theory for jamming and for the volume-fraction gradients}

Here I develop an approximate theory for the onset
of jamming in the $\Pefilm\gg 1$ regime. As in the previous
section I assume that the volume fraction of the big particles is so
small that it does not affect the small particles, which can be treated as a
one-component system.

\subsection{Fedorchenko and Chernov solution
for a diffusing ideal gas below a descending interface}

As in the earlier work of Sear and Warren \cite{sear17},
I will use the exact solution of Fedorchenko and Chernov
\cite{fedorchenko03, poon13,sear17}, for a diffusing
ideal gas in a film of infinite thickness ($H\to\infty$).
As discussed by Sear and
Warren \cite{sear17}, this solution can be used for
finite $H$, so long as the P\'{e}clet number
satisfies $\Pefilm \gg 1$.
After a short time $\tstar=1/\Pefilm$, an accumulation
zone is established below the interface. In that regime ($\tstar\Pefilm \gg 1$), the solution
of Fedorchenko and Chernov \cite{fedorchenko03}
(given in Appendix C of Sear and Warren \cite{sear17})
simplifies to
\begin{equation}
\phis(z, t)\approx\phiinit \left(1+
\Pefilm \tstar\exp\left[-\frac{|z-\zint|}{\Dsmall/\vev}\right]\right)
\label{phi_ideal}
\end{equation}
At the surface $z=\zint$, and we have
\begin{eqnarray}
\phis(\zint,\tstar)&=&\phiinit(1+\Pefilm \tstar)
\label{eq:phi_zint}
\end{eqnarray}
These equations only hold so long as $\phis<\phijam$,
beyond that jamming occurs.
Note that, see \Eqref{phi_ideal}, the  accumulation zone has
a constant width $\Dsmall/\vev$, and the maximum
concentration is at the interface and increases linearly with time.

\subsection{Jamming}

  %
  %

Jamming starts first at the surface as that
is where $\phis$ is highest. It starts when
the volume fraction there reaches the jamming volume fraction:
\begin{equation}
\phis(\zint,\tstarjam)=\phijam
\end{equation}
which defines the reduced evaporation time at which
jamming starts, $\tstarjam$.
If we use the simple approximation  of \Eqref{eq:phi_zint},
which is valid for $\Pefilm \tstar\gg 1$,
we obtain an estimate for the time at which jamming starts
\begin{equation}
\tstarjam\simeq\frac{\phijam/\phiinit-1}{\Pefilm}
\label{tstarjam}
\end{equation}

Drying films always jam.
Evaporation increases the
volume fraction until it hits
$\phijam$. However, to observe stratification, jamming
is not sufficient, we need the jammed layer to be preceded
by an accumulation zone where there is a steep concentration gradient.

This concentration gradient needs both a time of order $1/\Pefilm$
to become established and space to be established, a reduced height $z/H$
of $1/\Pefilm$ is enough.
Thus, we only have a jammed layer preceded by a steady-state
concentration profile unaffected by the bottom of the film,
when
\begin{equation}
\tstarjam< 1-1/\Pefilm
\end{equation}
Using, \Eqref{tstarjam}, this becomes
\begin{equation}
\phiinit>\frac{\phijam}{\Pefilm}
\label{pejamlimit}
\end{equation}
which must be satisfied for the jammed layer to appear
early enough.


Once a jammed layer has appeared,
we can use
simple mass conservation to obtain the steady-state
downward velocity of the jamming front, $\vjam$.
The flux of small colloidal particles into
the jammed region is just $\phiinit\vjam$, while the rate
of growth of the total volume fraction of small particles
in the jammed region is $\phijam(\vjam-\vev)$, where
$\vjam-\vev$ is the velocity at which the height of the jammed
region is increasing. If we just equate the flux
to the growth rate, and rearrange, we get
\begin{equation}
\vjam\simeq\frac{\vev}{1-\phiinit/\phijam}
\label{eq_vjam}
\end{equation}
The position of the
jamming front is then
\begin{equation}
\frac{\zjam(\tstar)}{H}\simeq 1-\tstar-(\tstar-\tstarjam)\left(\frac{\vjam}{\vev}-1\right)
~~~~ \tstar>\tstarjam
\end{equation}
From mass conservation, the jamming front reaches the bottom at time
\begin{equation}
\tstarend\simeq 1-\phiinit/\phijam
\end{equation}
defined by $\zjam(\tstarend)=0$, and we neglected a term of order $1/\Pefilm$.
The accumulation zone
will hit the bottom approximately $1/\Pefilm$ earlier.

Once a jammed layer has formed
the maximum gradient is at the front,
at $z=\zjam$. At steady state,
this maximum gradient is, see
Appendix \ref{app_jam},
\begin{equation}
\max\left(\frac{\partial\phis(z, t)}{\partial z}\right)=
\frac{\vjam(\phijam-\phiinit)}{\Dsmall}
\label{film_grad_jam}
\end{equation}

\subsubsection{Comparison of predicted gradients with experiment}

Using \Eqref{eq_vjam} the
maximum gradient can also
be written as
\begin{equation}
\max\left(\frac{\partial\phis(z, t)}{\partial z}\right)=\Pefilm H\frac{\phijam-\phiinit}{
1-\phiinit/\phijam}
\end{equation}
At constant initial film height
and initial volume fraction, my
simple model predicts that the
gradients in front of the jammed
region scale linearly with
$\Pefilm$. This is close to the
$\Pefilm^{0.8}$ dependence found
in experiments by
Ekanyake {\it et al.}\cite{Ekanayake:2009bs}.
Ekanyake {\it et al.}\cite{Ekanayake:2009bs}
varied $\Pefilm$ at constant
$H$ by increasing $\vev$, and
they report that the gradient
is measured below a \lq packed layer\rq,
so the experiments are in
comparable conditions to those
assumed by the model.
The difference between linear scaling, and
scaling as the power 0.8 is small,
so we have semiquantitative
agreement here.
Ekanyake {\it et al.}\cite{Ekanayake:2009bs}
compare with the model of
Routh and Zimmerman \cite{Routh:2004jz},
which predicts a $\Pefilm^{1/2}$ scaling.
As the experimental scaling lies in between
the two predictions, it is possible
that combining ideas from the
two models could give a model
in quantitative agreement with experiment,
but we leave this to future work.

\section{Diffusiophoresis in a drying film}

Having calculated the gradients in the volume fraction
of the small particles, I now determine the diffusiophoretic
velocities of the larger colloidal species.
The required expression for the diffusiophoretic velocity ${\bf U}$
in a suspension of much smaller particles
that are excluded from a layer of radius $\Rsmall$ from the
larger particle's surface is
\begin{equation}
{\bf U}(z, t)=-\frac{\Rsmall^2kT}{2\eta} \nabla\rhos
\label{dpvel1}
\end{equation}
where $\rhos$ is number density of
the smaller colloid, and $\eta$ is the viscosity.
This expression is well known \cite{And89,Bra11},
and was used by Sear and
Warren \cite{sear17}  for the Asakura-Oosawa ideal polymer
model \cite{asakura54}, although they were not the first to derive it \cite{And89,Bra11}.
Here we use this expression not for an ideal polymer but
for hard particles. The
two models differ only in the interactions between the small spheres.
Thus, \Eqref{dpvel1} will be a good approximation except at high
volume fractions of the small colloidal particles.

Using $\phis=(4\pi/3)\Rsmall^3\rhos$
and $\Dsmall=\kT/(6\pi\eta\Rsmall)$, we can
rewrite \Eqref{dpvel1} as
\begin{equation}
{\bf U}(z, t)=-\frac{9}{4}\Dsmall \nabla\phis
\label{dpvel}
\end{equation}
This is a general expression, we just need the gradient in the drying film.

Before jamming, the gradient is the derivative of \Eqref{phi_ideal}, which
gives
\begin{eqnarray}
U(z<\zint,\tstar<\tstarjam)&=&
\nonumber\\
\frac{9\phiinit\Pefilm\tstar\vev}{4}
&\exp&\left[-\frac{-|z-\zint|}{\Dsmall/\vev}\right]
\label{dpvel_film_nojam}
\end{eqnarray}
In the presence of a jammed layer,
the gradient
in the part of the film below
the jamming front is given by \Eqref{grad_postjam}.
So, the diffusiophoretic speed in the dilute
phase is
\begin{eqnarray}
U(z<\zjam,\tstar>\tstarjam)&=&
\nonumber\\
\frac{9(\phijam-\phiinit)\vjam}{4}
&\exp&\left[-\frac{-|z-\zjam|}{\Dsmall/\vjam}\right]
\label{dpvel_film}
\end{eqnarray}
which can also be written as
\begin{eqnarray}
U(z<\zjam,\tstar>\tstarjam)&=&
\nonumber\\
\frac{9\phijam\vev}{4}
&\exp&\left[-\frac{-|z-\zjam|}{\Dsmall/\vjam}\right]
\label{dpvel_film2}
\end{eqnarray}
if we use \Eqref{eq_vjam} for $\vjam$.
The maximum diffusiophoretic velocity in front of a jammed
layer is always simply $(9/4)\phijam\vev$, in our simple model.
This is because as $\phiinit$ increases, the increasing
$\vjam$ tends to increase the steepness of the gradient, but
this is exactly canceled by the decreasing total concentration
difference across the accumulation region: $\phijam-\phiinit$.

Following Sear and Warren \cite{sear17} I assume that the
diffusion of the large particles is negligible. Then the
dynamics of the large particles is just downward motion
at speed $\vbig(z, t)$, which is just diffusiophoretic motion
in the presence of a gradient of the small particles,
or motion at $\vev$ for particles at the interface or trapped
in the jammed state.
Thus, when there is a jammed layer, the speed of a large colloid is
\begin{equation}
\begin{array}{l}
\vbig(z, \tstar>\tstarjam)=\left\{\begin{array}{ll}
-\vev  & (z >\zjam) \\
-U  & (z < \zjam) \\
\end{array}\right.
\end{array}
\label{v_phor_film}
\end{equation}

\begin{figure}[tb!]
(a) \includegraphics[width=8.0cm]{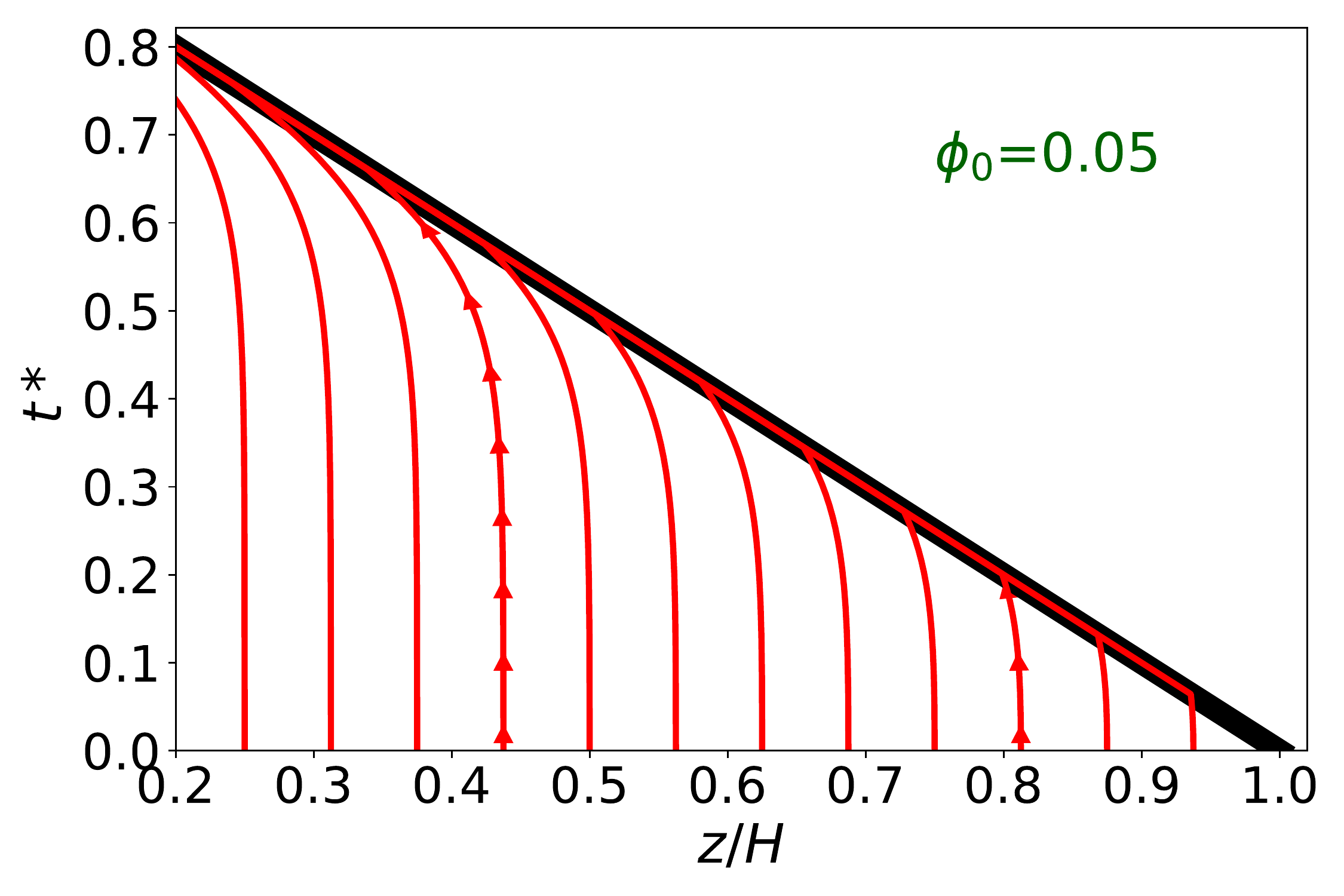}\\
(b) \includegraphics[width=8.0cm]{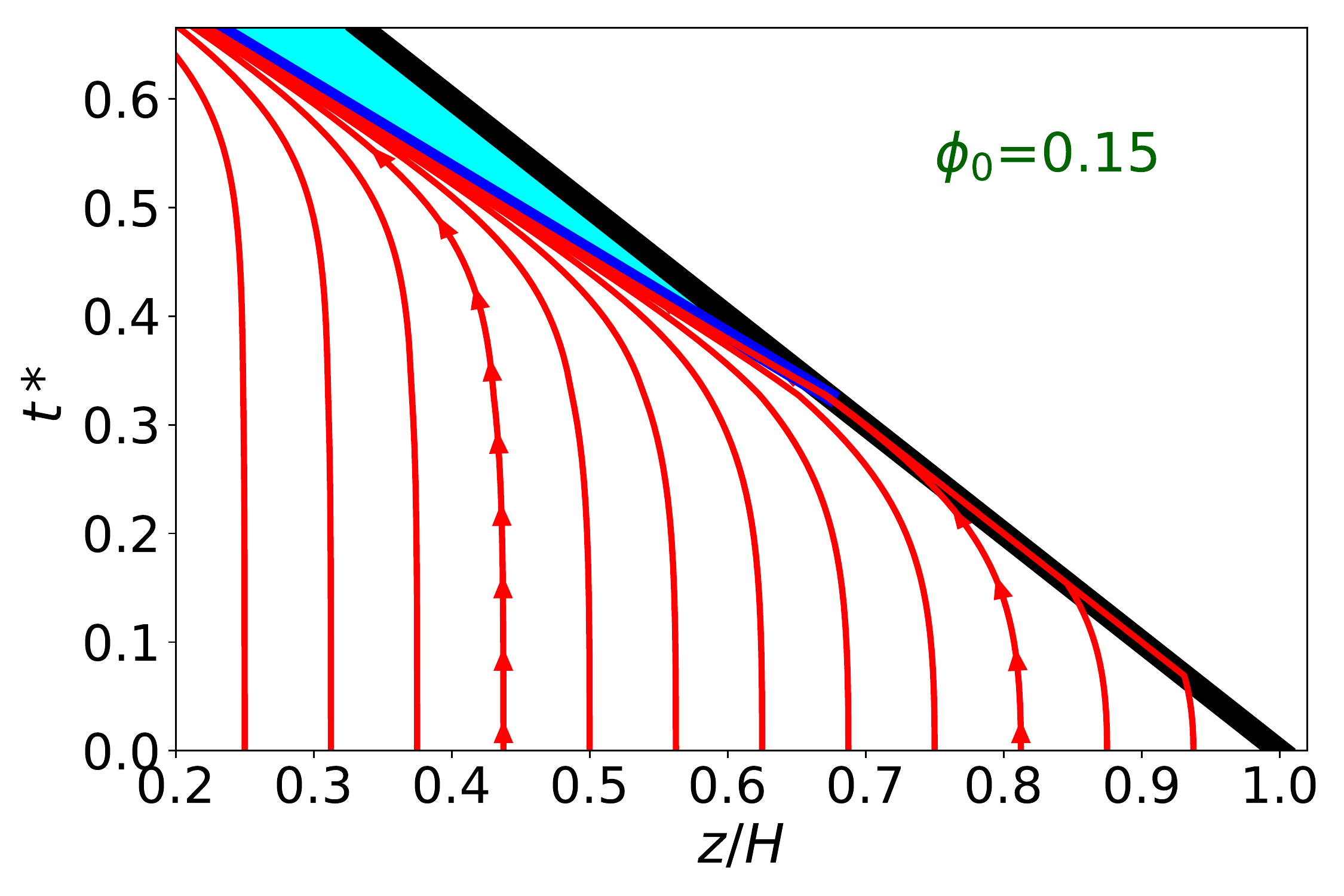}\\
(c) \includegraphics[width=8.0cm]{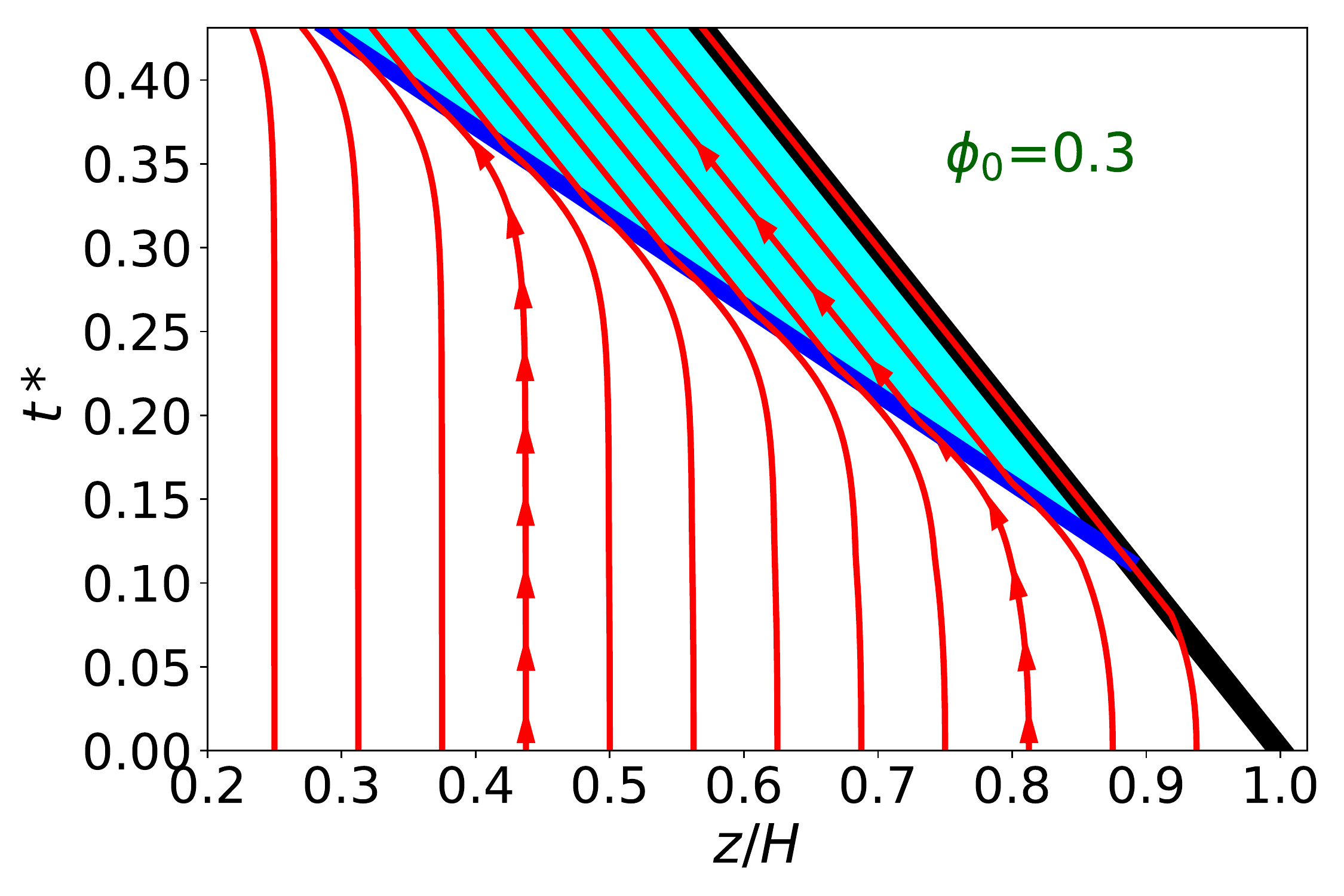}
 \caption{Trajectories $z_{\mathrm{big}}(t^∗)$ (red curves) of
 tracer large colloidal particles as a function of reduced time $t^*$.
(a), (b) and (c) are for initial volume fractions
$\phiinit=0.05$, $0.15$ and $0.30$, respectively.
In each panel two of the trajectories have arrows
to indicate the direction of the movement. The position of the top interface, $\zint$, is shown in black.
The jammed region is shaded in cyan,
and the yellow line is the jamming front at $\zjam$.
Calculations are for $\Pefilm = 10$ and
\label{separatrix}}
\end{figure}

\section{Results}

Now that I have expressions for both when jamming
occurs, and for the diffusiophoretic velocity,
I can make predictions for the behaviour of the large colloidal particles.
I neglect diffusion of the large colloidal particles. Then
the position of a large particle $z_{\mathrm{big}}$ is simply obtained by integrating
${\rm d}z_{\mathrm{big}}/{\rm d}t=\vbig$, with
the initial condition being the initial position of the
particle in the film.

In \Figref{separatrix},
I have plotted the trajectories during drying of a set
of particles with
equispaced initial
positions in the film. This is done for three values of the initial
volume fraction of the smaller colloid.
In \Figref{separatrix}(a) the film has a
small initial volume fraction of the small particles.
Then the volume fraction of small particles at the interface only becomes
large when drying is almost over, and the water/air interface is close
to the bottom surface. So no large gradients develop and
there is no stratification with a layer of small on top of a layer of big particles.

Note the convergence of the trajectories of the big particles at the top
interface, the slow moving big particles accumulate at the top.
By neglecting diffusion of the big particles we have effectively set their P\'{e}clet
number to be infinite. Trueman {\it et al.}\cite{trueman12,trueman12langmuir}
have developed models and present experimental data, for
the accumulation of big particles at the top of the drying film, due
to the large P\'{e}clet number of this species.
At very low concentrations of the small particles, interactions between the small and
big colloidal particles may be insignificant, and so the dominant
difference between the small and big particles is the much larger P\'{e}clet number
of the big particles.

In \Figref{separatrix}(b) the film has an intermediate
volume fraction of the small particles.
A jammed layer appears at $\tstar=0.33$, and so at an initial
height $z/H=0.67$. So when the jammed layer appears there is space underneath
it for a large concentration gradient to form.
This large gradient drives fast diffusiophoretic motion of the big particles, and
so the final film is stratified.
Note that just below the descending jamming front (yellow line)
there is strong curvature  of the trajectories (red)
away from the front.

It is worth noting that in this model
the large particles concentrate in a narrow
region in front of the descending
jammed region --- the red curves in
\Figref{separatrix}(b) converge on each other
and on the yellow line marking the descending front. Similar localisation of particles
due to diffusiophoresis is seen
in systems where diffusiophoresis is due to salt gradients \cite{SUS+16,ault17,shin17,shi16}.
There this convergence is called focusing.

Finally, in \Figref{separatrix}(c),
the film has a large initial
volume fraction of the small particles.
A jammed layer appears
at $\tstar=0.11$, and so at an initial
height $z/H=0.89$. So as at the intermediate volume fraction, \Figref{separatrix}(b),
a jamming layer forms with concentration gradients underneath it.
However, the diffusiophoretic velocity $U$ is too slow for the big particles
to outrun the descending jamming front, and big particles are incorporated into
the jammed layer. Note the red trajectories
that start in the unjammed region (white) but are incorporated in
the growing jammed region (cross the yellow line into the cyan region).

\begin{figure}[tb!]
\includegraphics[width=9.0cm]{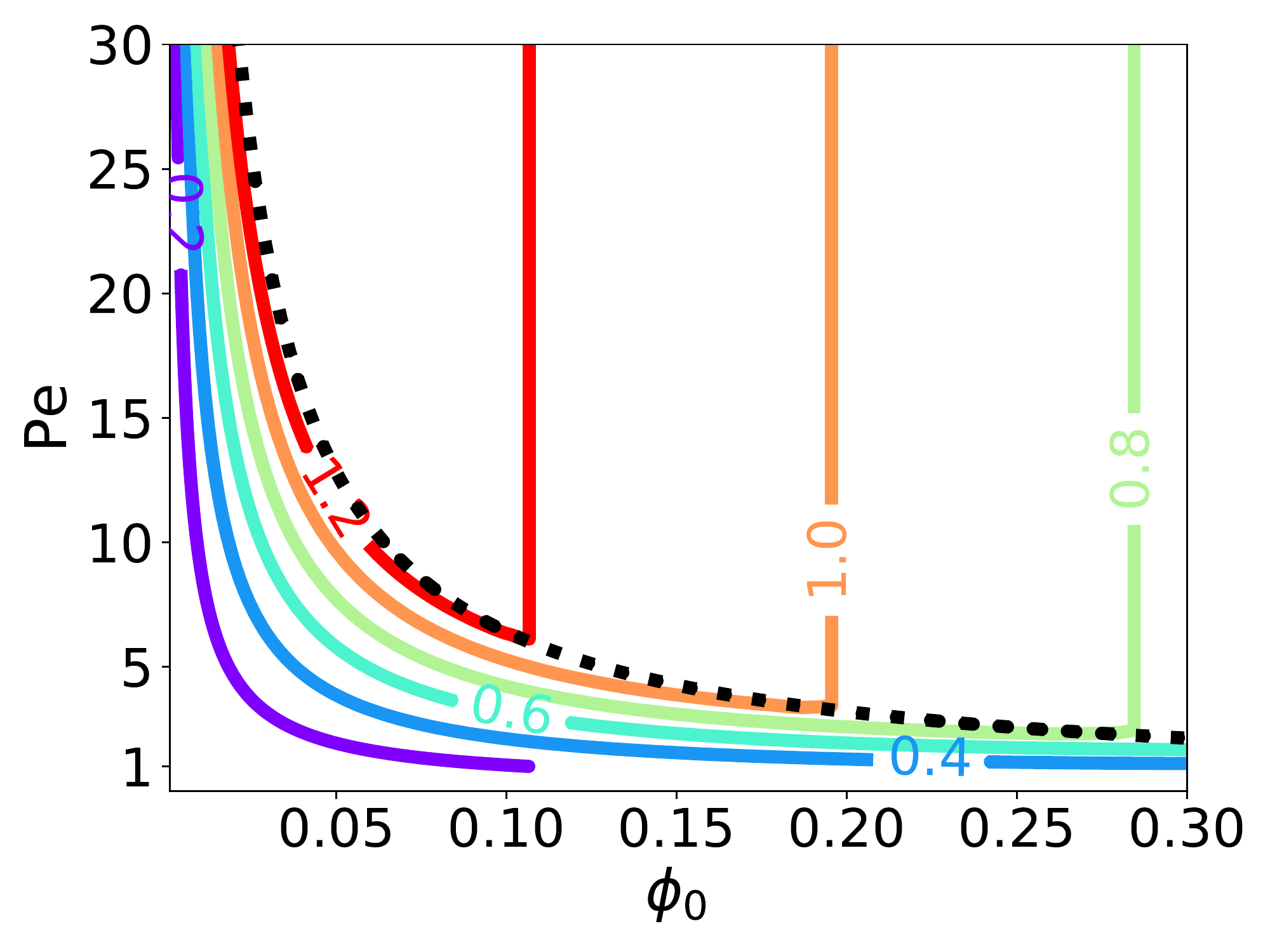}
 \caption{A contour plot of the ratio $\max(U)/\vjam$
 in the $\phiinit$-$\Pefilm$ plane. Superimposed
 on this is a black dotted curve, which is \Eqref{pejamlimit},
 and separates the regions of the plane where jamming
 occurs before the accumulation zone is limited by the bottom
 of the film (to right and above the curve), from the region
 where jamming only occurs when the interface is already close
 to the bottom of the film.
\label{contour}}
\end{figure}

\subsection{Region of the $\phiinit$--$\Pefilm$
plane where a jammed layer forms and excludes
the big particles}

In my simple model, stratification forms when a jammed
layer of the small particles forms and excludes the big particles.
Thus, there are two conditions that need
to be met for stratification:
1) a jammed layer must form early enough in drying so
that there is space ($\sim \Dsmall/\vjam$) below the jammed layer
for concentration gradients, and
2) the diffusiophoretic velocity due to these concentration gradients
must be fast enough
to push the big particles ahead of the jammed layer, $U>\vjam$.

Condition 1) is just \Eqref{pejamlimit}. For condition 2) we need
the maximum diffusiophoretic velocity.
Before jamming, as determined by \Eqref{pejamlimit},
the maximum is at the interface, see \Eqref{dpvel_film_nojam}.
When there is jamming, the maximum of $U$ is at $\zjam$, from \Eqref{dpvel_film}.
So,
\begin{equation}
\begin{array}{l}
\max(U)=\left\{\begin{array}{ll}
(9/4)\phiinit\Pefilm t^*\vev  & \tstar <\tstarjam \\
(9/4)(\phijam-\phiinit)\vjam & \tstar > \tstarjam \\
\end{array}\right.
\end{array}
\label{umax_film}
\end{equation}
Thus we can determine the value of the ratio $\max(U)/\vjam$ at all
values of $\phiinit$ and $\Pefilm$.
Figure \ref{contour} is a contour plot of the
ratio $\max(U)/\vjam$ (note that both $U$ and $\vjam$ depend on $\phiinit$).
The orange contour at $1.0$ separates the region where
$U$ is fast enough for stratification, from the region where it is too slow.
At its right-hand side the contour at 1.0 is vertical, i.e., is independent
of $\Pefilm$, because both the competing velocities
($U$ and $\vev$) are linear in $\Pefilm$.

We can determine this right-hand boundary of the stratified region by finding
where the ratio $\max(U)/\vjam=1$. Using \Eqref{umax_film} in the jammed region,
we then have
that $(9/4)(\phijam-\phiinit)=1$, or $\phiinit=\phijam-4/9=0.20$,
with $\phijam=0.64$.
When the initial volume fraction of the small particles
is greater than $0.20$, the diffusiophoretic velocity is too slow to push the big
particles ahead of the advancing jammed layer,
and stratification is impossible.

However, if the initial volume fraction of the small particles
is below $\phijam/\Pefilm$ then the jammed layer forms too late
in drying to drive stratification. The jammed layer only forms
when the accumulation zone of width $\Dsmall/\vev$ has already
reached the bottom. So stratification only
occurs for $\phiinit$ between $\phijam/\Pefilm$ and $0.20$.

\begin{figure}
\includegraphics[width=9.0cm]{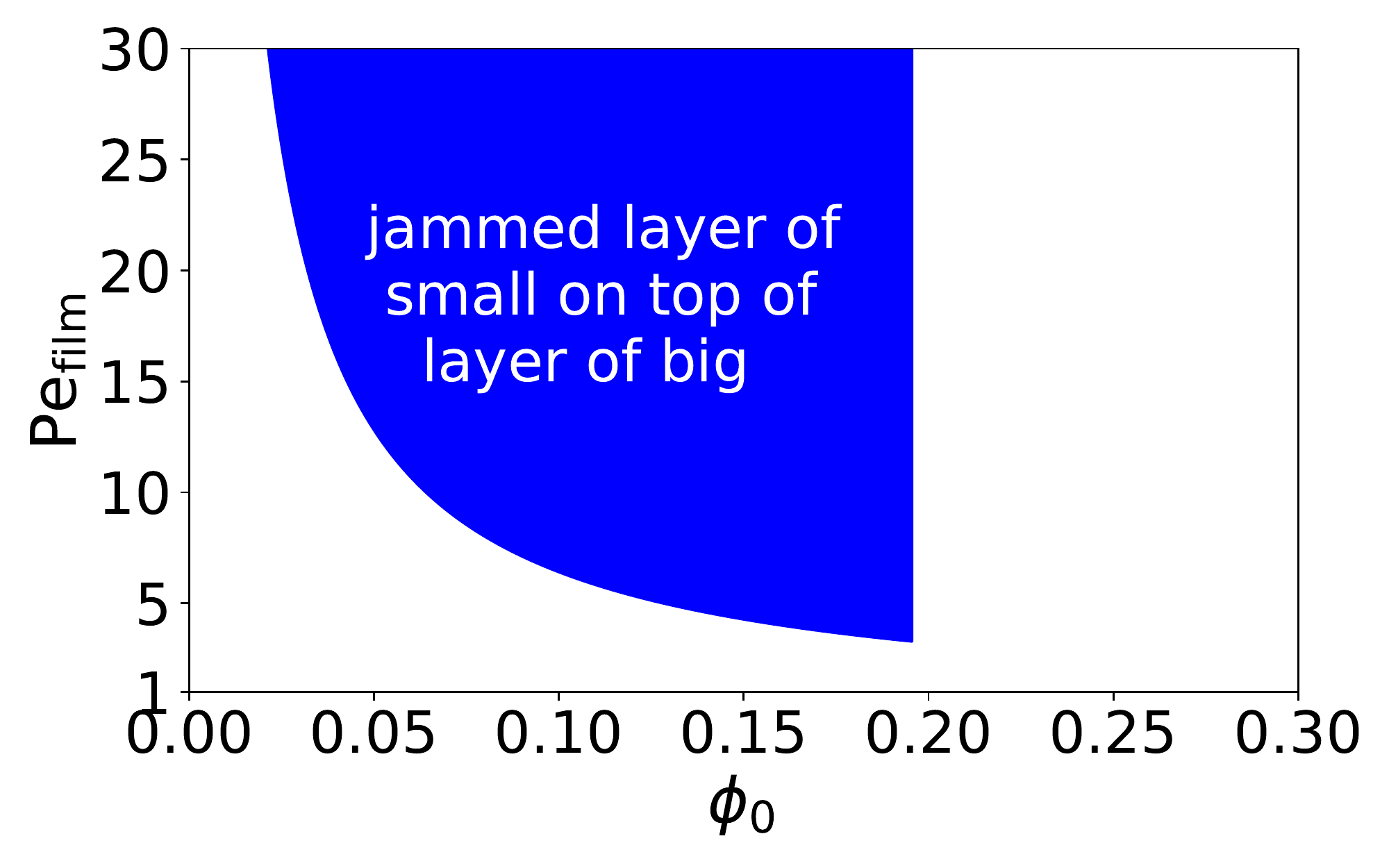}
\caption{Plot of the $\phiinit$-$\Pefilm$ plane,
with the region where there is stratification shown in blue.
This is the region which satisfies \Eqref{pejamlimit},
and where $\max(U)>\vjam$.
\label{phi_peclet}}
\end{figure}

In \Figref{phi_peclet}, I show the
$\phiinit$-$\Pefilm$-plane, and have shaded in blue the region where stratification occurs.
This figure follows a similar plot made by Zhou \etal\ \cite{zhou17} for
their model. Sear and Warren \cite{sear17} show this type
of plot, for a model without jamming, and
Makepeace {\it et al.} \cite{makepeace17} and Liu {\it et al.}\cite{liu18} both
plot experimental data in this way.
See the review
of Schulz and Keddie \cite{schulz18}
for earlier experimental work including
the conditions where stratified and non-stratified films
have been observed.

\section{Conclusion}

As we can see in \Figref{phi_peclet}, drying films stratify
over a range of initial volume fractions of the small colloid.
The lower limit to stratification decreases
as the P\'{e}clet number increases. This lower limit
 is set by the fact
that below it, there are so few small particles that jamming only
occurs when the water/air interface is already close to the bottom
of the film ($\zint/H<1-1/\Pefilm$).
The upper limit is set by the fact that as the
concentration of the small particles increases, the speed of advance
of the jamming front increases but the diffusiophoretic velocity does not.
So at volume fractions $\phiinit>0.20$, the big particles no longer move fast enough to outrun the
advancing jamming layer.

The prediction that large initial
concentrations of small particles do not
result in stratification is
consistent with the results of both Mart\'{i}n-Fabiani {\it et al.} \cite{martin16},
and Makepeace {\it et al.} \cite{makepeace17}, and with the results
surveyed in Fig.~10 of the
review of Schulz and Keddie \cite{schulz18}.
Schulz and Keddie plot the results of many experiments
on drying films of colloidal films,
and find only few stratify at volume
fractions of the smaller species above 0.2,
and none above a volume fraction of
approximately 0.3, although it should be noted that there is a
little data in that region. Schulz and Keddie also find that
most systems where the initial
volume fraction of the smaller particles
is much less than 0.1, also do not stratify.

With the exception of the work of Cheng and coworkers \cite{Cheng:2013cp,Cheng:2016ep},
computer simulation
studies \cite{fortini16,fortini17,makepeace17,martin16,HNP17a,HNP17b}
have studied systems with implicit not explicit solvent.
As discussed in detail by Sear
and Warren \cite{sear17},
computer simulations of models without explicit solvent, neglect
solvent flow effects and so
overpredict stratification.
So, it is only because our model includes solvent-flow effects,
that it is able to make
the prediction that
stratification only occurs over a limited
range of volume fractions of the smaller species.
Simulations with explicit solvent \cite{Cheng:2013cp,Cheng:2016ep}
are very challenging computations, and so are forced to study systems
at larger evaporation rates and thinner films, than studied in experiment.
This makes it difficult to directly compare the interesting results of
simulations with explicit solvent, with experiment.

I would like to end by making a few remarks
on future work. We now have a number of experimental studies with data
on the final dry films \cite{schulz18}. We also have models for
the dynamics during drying that make
clear predictions. However, there is still a lot of work to do
before we can confidently say we understand and can rationally engineer
drying films containing colloidal mixtures.

Our current models are all incomplete and make approximations.
Here I assumed that the volume fraction of the big particles was so
small that I could neglect interactions between big particles, and
also that the size ratio $\Rbig/\Rsmall\gg 1$.
In addition, not all possible behaviour has been considered.
For example, the coupling of stratification and crystallisation has not been
considered.
Mixtures can often only crystallise with
difficulty \cite{zhang14,pusey09,rios17,dijkstra99}, however, stratification
demixes mixtures and creates a layer of almost
pure small particles, which may then go to crystallise.
Thus stratification may allow mixtures that
would otherwise remain amorphous to crystallise.
Future modelling work could consider this. It could also
consider the effect of varying the size ratio,
$\Rbig/\Rsmall$, by using available expressions for $U$ as
a function of size ratio \cite{anderson91}.

Further experiments are also needed.
Most experimental studies report only on the final dry film,
although the work of Ekanayake {\it et al.}\cite{Ekanayake:2009bs}, and that
of Cardinal {\it et al.}\cite{Cardinal:2010be} are exceptions.
So, we have little data on the dynamics of colloidal mixtures during drying.
To fully understand the processes during drying that lead to
stratification, future experimental work will need to study
particle dynamics during the drying process.

\begin{acknowledgments}

I would like
to thank Andrea Fortini, Joseph Keddie and Patrick Warren for many helpful conversations.

\end{acknowledgments}

\begin{widetext}

\appendix

\section{Concentration gradient in the small particles below a jammed layer}
\label{app_jam}

For an ideal gas in front of an advancing jammed front at
position $\zjam$, the decay to the uniform value
is exponential, with a characteristic width $\Dsmall/\vjam$, as shown
by Okuzono {\it et al.}~\cite{okuzono06} (see their Eq.~(18)).
The profile is then given by
\begin{equation}
\phis(z, t)\approx\left\{\begin{array}{cc}
\phijam & \zjam < z < \zint  \\
\phiinit+(\phijam-\phiinit)\exp\left[-\frac{-|z-\zjam|}{\Dsmall/\vjam}\right]
& z< \zjam  \\
\end{array} \right.
\end{equation}
Note that below the descending interface there is an accumulation zone, where
the volume fraction $\phis>\phiinit$,. This zone is of constant width $\Dsmall/\vjam$.
The gradient in volume fraction of small particles is then
\begin{equation}
\frac{\partial \phis(z, t)}{\partial z}
\approx\left\{\begin{array}{cc}
0 & \zjam < z < \zint \\
\frac{(\phijam-\phiinit)\vjam}{\Dsmall}
\exp\left[-\frac{-|z-\zjam|}{\Dsmall/\vjam}\right]
& z< \zjam  \\
\end{array} \right.
\label{grad_postjam}
\end{equation}
The maximum in the gradient
is at the advancing jamming front, i.e., at $\zjam$. Putting $z=\zjam$
in \Eqref{grad_postjam} yields \Eqref{film_grad_jam}.
Equation (\ref{film_grad_jam})
applies so long as the accumulation zone, of width $\Dsmall/\vev$,
that precedes the jamming front, does not hit the bottom of the film, i.e.,
so long as $\zjam>\Dsmall/\vjam$.
The solution for this system in the $H\to\infty$ limit
is given by Landau \cite{landau50}.



\end{widetext}


%

\end{document}